\documentclass[pdftex,twocolumn,epjc3]{svjour3}          

\usepackage{graphicx}
\usepackage{dcolumn}
\usepackage{bm}
\usepackage{amssymb}
\usepackage{enumerate}
\usepackage{lineno}
\usepackage{multirow} 
\usepackage{slashed}

\RequirePackage{graphicx}
\RequirePackage{mathptmx}   
\RequirePackage{flushend}
\RequirePackage[numbers,sort&compress]{natbib}
\RequirePackage[colorlinks,citecolor=blue,urlcolor=blue,linkcolor=blue]{hyperref}

\usepackage{graphicx}
\usepackage{dcolumn}
\usepackage{bm}
\usepackage{amssymb}
\usepackage{enumerate}
\usepackage{lineno}
\usepackage{textalpha} 
\usepackage{doi} 

\usepackage{lineno}

\newcommand\gmu{(g_{\mu}-2)/2}

\newcommand\ee{e^+e^-}

\newcommand\ainv{A'\to invisible}
\newcommand\chidecay{\chi_2 \to \chi_1 e^+ e^-}
\newcommand\asemi{\gp \to \chi_1 \chi_2 (\chi_2 \to \chi_1 e^+ e^-)}
\newcommand\inv{ \chi\overline{\chi}}
\newcommand\gp{ A'}

\newcommand\pair{e^+e^-}

\newcommand\dimu{\mu^+ \mu^-}

\def\address{\@ifstar{\address@star}%
  {\@ifnextchar[{\address@optarg}{\address@noptarg}}}

\begin{document}

\title{
Probing the explanation of the muon (g-2) anomaly and thermal light dark matter with the semi-visible dark photon channel
}



\subtitle{NA64 Collaboration}
\author{
C.~Cazzaniga\thanksref{addrETH,pa} \and
P.~Odagiu\thanksref{addrETH,addrEPFL,pa} \and
E.~Depero\thanksref{addrETH,pa} \and
L.~Molina~Bueno\thanksref{addrETH,addrCSIC,pa} \and
Yu.~M.~Andreev\thanksref{addrINR} \and
D.~Banerjee\thanksref{addrCERN,addrUIL} \and
J.~Bernhard\thanksref{addrCERN} \and
V.~E.~Burtsev\thanksref{addrJINR} \and
N.~Charitonidis\thanksref{addrCERN} \and
A.~G.~Chumakov\thanksref{addrTGPU,addrTPU} \and
D.~Cooke\thanksref{addrUCL} \and
P.~Crivelli\thanksref{addrETH,pc} \and
A.~V.~Dermenev\thanksref{addrINR} \and
S.~V.~Donskov\thanksref{addrIHEP} \and
R.~R.~Dusaev\thanksref{addrTPU} \and
T.~Enik\thanksref{addrJINR} \and
A.~Feshchenko\thanksref{addrJINR} \and
V.~N.~Frolov\thanksref{addrJINR} \and
A.~Gardikiotis\thanksref{addrUPT} \and
S.~G.~Gerassimov\thanksref{addrTUM,addrLPI} \and
S.~Girod\thanksref{addrCERN} \and
S.~N.~Gninenko\thanksref{addrINR} \and
M.~H\"osgen\thanksref{addrUBN} \and
V.~A.~Kachanov\thanksref{addrIHEP} \and
A.~E.~Karneyeu\thanksref{addrINR} \and
G.~Kekelidze\thanksref{addrJINR} \and
B.~Ketzer\thanksref{addrUBN} \and
D.~V.~Kirpichnikov\thanksref{addrINR} \and
M.~M.~Kirsanov\thanksref{addrINR} \and
V.~N.~Kolosov\thanksref{addrIHEP} \and
I.~V.~Konorov\thanksref{addrTUM,addrLPI} \and
S.~G.~Kovalenko\thanksref{addrUNAB,addrSAPHIR} \and
V.~A.~Kramarenko\thanksref{addrJINR,addrMSU} \and
L.~V.~Kravchuk\thanksref{addrINR} \and
N.~V.~Krasnikov\thanksref{addrJINR,addrINR} \and
S.~V.~Kuleshov\thanksref{addrUNAB,addrSAPHIR} \and
V.~E.~Lyubovitskij\thanksref{addrTGPU,addrUSM,addrTPU,addrSAPHIR} \and
V.~Lysan\thanksref{addrJINR} \and
V.~A.~Matveev\thanksref{addrJINR} \and
Yu.~V.~Mikhailov\thanksref{addrIHEP} \and
D.~V.~Peshekhonov\thanksref{addrJINR} \and
V.~A.~Polyakov\thanksref{addrIHEP} \and
B.~Radics\thanksref{addrETH} \and
R.~Rojas\thanksref{addrUSM} \and
A.~Rubbia\thanksref{addrETH} \and
V.~D.~Samoylenko\thanksref{addrIHEP} \and
D.~Shchukin\thanksref{addrLPI} \and
H.~Sieber\thanksref{addrETH} \and
V.~O.~Tikhomirov\thanksref{addrLPI} \and
I.V.~Tlisova\thanksref{addrINR} \and
D.~A.~Tlisov\thanksref{dt,addrINR} \and
A.~N.~Toropin\thanksref{addrINR} \and
A.~Yu.~Trifonov\thanksref{dt,addrTGPU,addrTPU} \and
B.~I.~Vasilishin\thanksref{addrTPU} \and
G.~Vasquez Arenas\thanksref{addrUSM} \and
P.~V.~Volkov\thanksref{addrJINR,addrMSU} \and
V.~Yu.~Volkov\thanksref{addrMSU} \and
P.~Ulloa\thanksref{addrUNAB}
}

\thankstext{pa}{These first authors contributed equally to this article} 
\thankstext{pc}{Corresponding author, e-mail: Paolo.Crivelli@cern.ch} 
\thankstext{dt}{Deceased}

\institute{
ETH Z\"urich Institute for Particle Physics and Astrophysics,
CH-8093 Z\"urich, Switzerland\label{addrETH}                                    
\and
Laboratoire de Physique des Hautes Energies (LPHE),
École polytechnique fédérale de Lausanne (EPFL)
BSP Cubotron, CH-1015 Lausanne, Switzerland\label{addrEPFL}                     
\and 
Instituto de Fisica Corpuscular (CSIC/UV),
Carrer del Catedrátic José Beltrán Martinez, 2, 46980 Paterna, Valencia\label{addrCSIC} 
\and 
Institute for Nuclear Research, 117312 Moscow, Russia\label{addrINR}            
\and 
CERN, EN-EA, 1211 Geneva 23, Switzerland\label{addrCERN}                        
\and
University of Illinois at Urbana Champaign, Urbana,
61801-3080 Illinois, USA\label{addrUIL}                                         
\and 
Joint Institute for Nuclear Research, 141980 Dubna, Russia\label{addrJINR}      
\and 
Tomsk State Pedagogical University, 634061 Tomsk, Russia\label{addrTGPU}        
\and 
UCL Departement of Physics and Astronomy, University College London,             
Gower St. London WC1E 6BT, United Kingdom\label{addrUCL}                        
\and
State Scientific Center of the Russian Federation Institute for High Energy 
Physics of National Research Center 'Kurchatov Institute' (IHEP), 
142281 Protvino, Russia\label{addrIHEP}                                         
\and
Tomsk Polytechnic University, 634050 Tomsk, Russia\label{addrTPU}               
\and 
Physics Department, University of Patras, 265 04 Patras, 
Greece\label{addrUPT}                                                           
\and 
Technische Universit\"at M\"unchen, Physik  Department, 85748 Garching, 
Germany\label{addrTUM}                                                          
\and  
P.N. Lebedev Physical Institute, 119 991 Moscow, Russia\label{addrLPI}          
\and  
Universit\"at Bonn, Helmholtz-Institut f\"ur Strahlen-und Kernphysik, 
53115 Bonn, Germany\label{addrUBN}                                              
\and 
Departamento de Ciencias F\'{i}sicas, Universidad Andres Bello, Sazi\'{e} 2212, 
Piso 7, Santiago, Chile\label{addrUNAB}                                         
\and
Millennium Institute for Subatomic Physics at the High-energy frontier (SAPHIR),
ICN2019\_044, ANID, Chile\label{addrSAPHIR}
\and 
Skobeltsyn Institute of Nuclear Physics, Lomonosov Moscow State University, 
119991  Moscow, Russia\label{addrMSU}                                           
\and 
Universidad T\'{e}cnica Federico Santa Mar\'{i}a, 2390123 Valpara\'{i}so, 
Chile \label{addrUSM}                                                           
}

\maketitle

\begin{abstract}
We report the results of a search for a new vector boson ($\gp$) decaying into two dark matter particles $\chi_1 \chi_2$ of different mass. The heavier $\chi_2$ particle subsequently decays to $\chi_1$ and $\gp \to \pair$. For a sufficiently large mass splitting, this model can explain in terms of new physics the recently confirmed discrepancy observed in the muon anomalous magnetic moment at Fermilab. Remarkably, it also predicts the observed yield of thermal dark matter relic abundance. A detailed Monte-Carlo simulation was used to determine the signal yield and detection efficiency for this channel in the NA64 setup. The results were obtained re-analyzing the previous NA64 searches for an invisible decay $A'\to \inv$ and axion-like or pseudo-scalar particles $a \to \gamma \gamma$. With this method, we exclude a significant portion of the parameter space justifying the muon g-2 anomaly and being compatible with the observed dark matter relic density for $A'$ masses from 2$m_e$ up to 390 MeV and mixing parameter $\epsilon$ between $3\times10^{-5}$ and $2\times10^{-2}$.
\end{abstract}

Despite its great success, the Standard Model (SM) does not provide a complete description of nature. For example it cannot explain the origin of dark matter, the neutrino masses and the baryon asymmetry problem. Furthermore, interesting discrepancies between some SM predictions and measurements have been observed. These include the LHCb results challenging lepton universality \cite{Aaij:2021vac} and the long standing discrepancy of the muon anomalous magnetic moment $a_{\mu}=(g_{\mu}-2)/2$ \cite{Bennett:2004pv} which was recently confirmed \cite{PhysRevLett.126.141801}. The combination of the Brookhaven and Fermilab muon g-2 results compared to the latest theoretical calculations using dispersion relations\cite{AOYAMA20201} leads to a discrepancy of $\sim 4.2\sigma$:
$$ \Delta a_{\mu} \equiv a_{\mu}(exp) - a_{\mu}(th) =(251 \pm 59)\cdot 10^{-11}.$$
It should be noted that when compared to the latest QCD lattice calculations this is reduced to about 1.4$\sigma$ \cite{Borsanyi:2020mff}. In order to help elucidate the origin of this difference, a new experiment aiming to measure the contribution of hadronic corrections is being prepared at CERN \cite{Abbiendi:2020sxw}.

\noindent In terms of new physics, among many interesting proposals, a way to explain this discrepancy is to introduce a 1-loop correction involving a $U(1)_{D}$ dark sector massive gauge field $A'$ to the QED 3-point vertex \cite{Pospelov:2008zw}. The Dark Photon $A'$ can couple to both charged Dark Matter (DM) fields $\chi_i$ with coupling strength $g_D$, and to SM leptons via kinetic mixing $\epsilon$ with the SM photon field $A$.
Considering models with a diagonal $\gp$ coupling to DM and SM fields,
two decay modes are possible,  $\gp \to \pair$ (visible mode) \cite{visible-2018-analysis} and  $\gp \to \inv $ (invisible mode)\cite{Banerjee_2019}. These were excluded as explanations of the $a_{\mu}$ anomaly by the combined results of NA64 and BaBar \cite{babar, na64-prd, Banerjee_2019} (for $\gp \to \inv$) and NA48 \cite{na48} (for $\gp \to \ee$).
In addition, the prospected sensitivity for NA64 running in muon mode to probe the $(g_\mu-2)/2$ anomaly 
was estimated recently \cite{Kirpichnikov:2020tcf} in the anomaly free $L_\mu-L_\tau$ gauge extension of the SM.

\noindent In this work we consider an alternative model for $\gp$ decay. This was initially proposed to recover an explanation of the $g_{\mu}-2$ discrepancy still within the Dark Photon paradigm \cite{Izaguirre:2017bqb,Mohlabeng:2019vrz}. This model is characterised by the $\gp$ decaying into a heavier $\chi_2$ and a lighter $\chi_1$ Dark Matter states. While $\chi_1$ is a non-interacting stable state which determines the DM relic abundance, $\chi_2$ is unstable and de-excites by emitting a $\chi_1$ and an off-shell $\gp^{\ast}$, that subsequently decays into an electron-positron pair. Thus, a new semi-visible decay mode combining characteristics of both the visible and invisible decay signatures emerges. It is very remarkable that such a model can potentially explain both $\gmu$ and the Dark Matter thermal relic abundance for 300 MeV$\lesssim m_{\gp} \lesssim$ 1 GeV, thus making it of great phenomenological interest \cite{Izaguirre:2017bqb,Mohlabeng:2019vrz}.


In this study, we focus on a Dark Matter model that extends the SM symmetry group with a dark sector $U(1)_D$, which is spontaneously broken by a dark Higgs field $h_D$ \cite{Mohlabeng:2019vrz}. The gauge mixing with the SM photon via the term $-\epsilon F^{\mu \nu}[A']F_{\mu \nu}[A]$, governed by the parameter $\epsilon$, generates a massive Dark Photon after spontaneous symmetry breaking.

\noindent The diagonalisation of the mixed gauge interaction \cite{Feldman:2007wj} allows the removal of $\epsilon$ from the pure mixed-gauge Lagrangian. However, a new coupling appears between $\gp$ and the SM electromagnetic current, with interaction strength~$\epsilon e$. The novel interaction term is the main phenomenological feature of the model: the possibility to have SM final states produced via kinetic mixing of $\gp$ with $A$. Nevertheless, the unique feature of the semi-visible model, that distinguishes it from the invisible and visible channels, is found in the Yukawa dark sector. In the unbroken theory, we start from a pseudo-Dirac field $\Psi$ charged under $U(1)_D$. The chiral projections of $\Psi$ can then be coupled with $h_D$ to produce Dirac and Majorana mass terms, after spontaneous symmetry breaking \cite{Mohlabeng:2019vrz}. 

\noindent In this pseudo-Dirac scenario, the right and left Majorana masses are the same and are strongly suppressed relative to the Dirac mass $M_D$. The diagonalisation of the mass matrix results in two eigenstates: a lighter stable $\chi_1$ and a heavier $\chi_2$ DM particle, with mass difference $\Delta \equiv m_{\chi_2} - m_{\chi_1}$ \cite{LDMXproposal}. 

\begin{figure}
\includegraphics[scale=0.50]{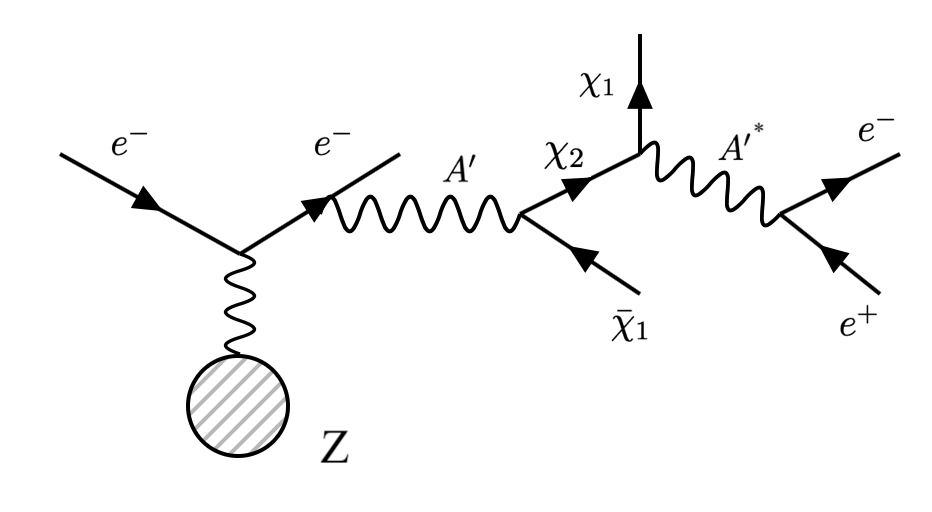}
\caption{\label{fig:epsart} Production of $\gp$ and subsequent semi-visible decay chain of a Dark Photon, $e^- Z \to e^-Z\gp;\asemi$.}
\end{figure}

\noindent The effective Lagrangian for the semi-visible model is:
\begin{eqnarray}\label{eq:lagr} 
\mathcal{L} = & \mathcal{L}_{SM} - \frac{1}{4}F^{\mu \nu}[A']F_{\mu \nu}[A']+\frac{1}{2}m_{A'}^2A'^2+ \epsilon e\bar{\psi_e}\slashed{A}'\psi_e\nonumber \\&
+\sum_i\bar{\chi}_{i}(\slashed\partial - m_{\chi_i})\chi_i + (g_D\bar{\chi}_{2}\slashed{A}'\chi_{1} + h.c).   
\end{eqnarray}
In Eq.\ref{eq:lagr}, the coupling to muons is neglected since the di-muon production threshold $\Delta = 2m_{\mu}$ is not relevant for the sub-GeV mass range of $\gp$ probed in this study. 
The absence of elastic diagonal interaction terms $\sum_i\bar{\chi}_{i}\slashed{A}'\chi_i$ derives from the choice of a pseudo-Dirac field, where only off-diagonal terms are permitted. This allows us to neglect the invisible and visible $\gp$ decay modes, where conversely only diagonal terms are present. A sketch of the dominating decay chain is shown in Fig.\,\ref{fig:epsart}. The Dark Photon $A'$ decays promptly in a lighter $\chi_1$ and a heavier $\chi_2$ via the mentioned inelastic interaction, followed by the subsequent decay $\chi_2 \to \chi_1 e^{+}e^{-}$.

\noindent The width of the process was calculated at leading order. A numerical approach was used to compute the 3-body decay phase-space, implemented in the module MadWidth of MadGraph5aMC@NLO \cite{Alwall:2014bza}. Thus, a correction is obtained to the previous analytical approximations (valid for $m_{A^{'}} \gg m_{\chi1} \gg m_e$) of the $\Gamma (\chi_2 \to \chi_1 \pair)$ formula from Refs.\,\cite{Mohlabeng:2019vrz,Izaguirre:2015zva}. The newly attained formula reads
\begin{eqnarray}\label{eq:width} 
\Gamma (\chi_2 \to \chi_1 \pair) \simeq K \frac{4 \epsilon^2 \alpha_{EM}\alpha_{D}\Delta^5}{15 \pi m_{A'}^4},
\end{eqnarray}
where $K \simeq 0.640 \pm 0.004$ is the correction factor estimated using both Madgraph and CalcHep \cite{Belyaev:2012qa}, found to be in good agreement. The quoted uncertainty on the $K$ factor takes into account the difference between Madgraph and CalcHep and the slight dependence of $K$ on $m_A'$. The K-factor was found to be basically insensitive to the other parameters of the model. 

An upper bound for the dark sector coupling constant $\alpha_D$  can be found by requiring the absence of a Landau pole for the effective coupling constant $\bar{\alpha}_D(\mu)$ up to an energy scale $\Lambda \sim 1$\,TeV: $\alpha_D \lesssim 0.2$ \cite{COMB,DSR}. In this study, a benchmark value of $\alpha_D = 0.1$ is used, compatible with other literature \cite{Izaguirre:2017bqb,Mohlabeng:2019vrz}. Nevertheless, a discussion on the implications of different $\alpha_D$ choices in our results will be provided. Furthermore, the resonance of the thermal averaged non-relativistic co-annihilation DM cross section $ \langle \sigma_{an}v_{rel} \rangle (\chi_1 \chi_2 \to \pair)$ present at $ m_{A'} \sim 2 m_{\chi_1}$ \cite{Izaguirre:2015zva,LDMXproposal} can be avoided by setting as benchmark $m_{A'} = 3 \cdot m_{\chi_1}$ such as in \cite{Mohlabeng:2019vrz,LDMXproposal,Izaguirre:2015zva}. Finally, the parameter $\Delta$ has kinematic limits $\Delta < m_{\gp} - 2 \cdot m_{\chi_1}$ and $\Delta > 2 m_e$. A relatively large mass splitting $\Delta/m_{\chi_1} = 0.4$ is chosen in this study, as strong bounds for lower $\Delta$ already exist as explanation of $\gmu$ by BABAR and E137\cite{babar,Mohlabeng:2019vrz,Izaguirre:2017bqb}. A complete discussion of the achievable $\Delta$ range, up to the limit $\Delta/m_{\chi_1} \simeq 1$, is provided below.

\begin{figure*}[tbh!!]
\includegraphics[width=.8\textwidth]{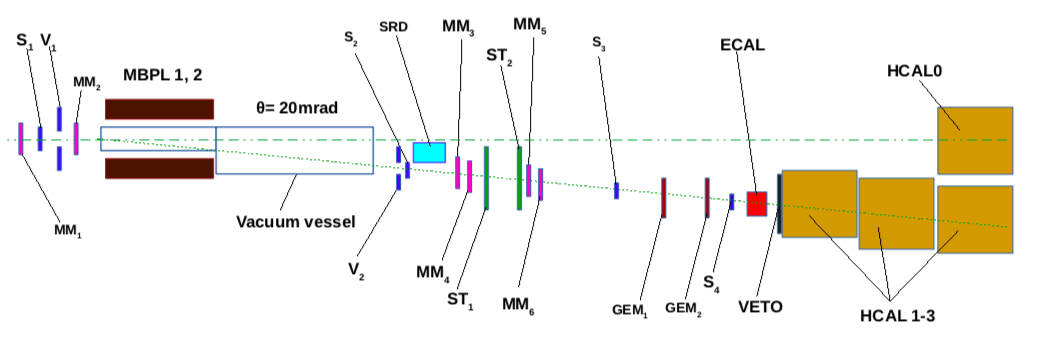}%
\vskip-0.cm{\caption{A schematic view of the NA64 invisible mode setup used in 2018 \label{fig:invis-setup}}}
\end{figure*}

In this work, we present a direct search for the $\gp$ semi-visible signature using the NA64 experiment located at CERN SPS. The Dark Photons are produced in the process $e^-Z \rightarrow e^-Z A'$ as 100 GeV electrons coming from the H4 beamline scatter inside the NA64 electromagnetic calorimeter (ECAL). The production mechanism is identical to the one described in Ref.\,\cite{Banerjee_2019}. The setup is schematically shown in Fig.\,\ref{fig:invis-setup}.
The experiment uses a set of scintillator and veto counters, a magnet spectrometer consisting of two dipole magnets, and a set of tracking detectors (six micromegas chambers \cite{na64-micromegas}, three straw detectors \cite{Volkov:2019qhb} and two GEMs \cite{gem}) to define the incoming $e^-$ beam. A synchrotron radiation detector (SRD) is used to suppress the hadron contamination in the beam. The electrons are absorbed in a lead-scintillator sandwich Shashlick-type ECAL target of 40 radiation lengths. Downstream from the ECAL, a large high-efficiency VETO counter and three iron hadronic calorimeters (HCALs) complete the setup. An additional HCAL module is placed along the unbent beam path to further suppress background from upstream $e^-$ interactions. Further details about the setup can be found in Refs.\,\cite{Banerjee_2019, Banerjee-ALP-2020, Dusaev:2020gxi}.

\begin{figure}[tbh!!]
\includegraphics[width=.45\textwidth]{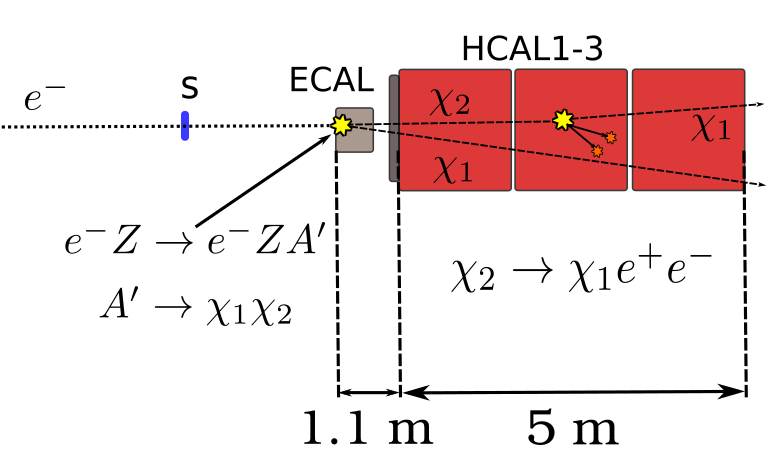}
\vskip-0.cm{\caption{A schematic view of an event $\asemi$ from a $\gp$ 
produced after a 100 GeV $e^-$ scatters off in the active dump, $e^- Z \to e^- Z \gp$. The $\chi_2$ particle decaying within HCAL2 corresponds to the S1 signature (see text for more details).\label{fig:semivis-sig}}}
\end{figure} 

The $\gp$ is produced in the target via Dark-Bremsstrahlung \cite{jdb} and decays promptly into $\chi_1 \chi_2$. The long-lived $\chi_2$ travels through the VETO and HCAL1, which acts as a veto to reject particles leaking from the target, before decaying through $\chidecay$. The result of this decay would be observed in the experiment through two characteristic signatures. The first (S1), is identified by the presence a significant energy deposition in HCAL2 or HCAL3, as the result of the decay to $\ee$ (see Fig.\ref{fig:semivis-sig}). 
In the second signature (S2), $\chi_2$ decays outside the fiducial volume limited by the last HCAL module, traveling a distance $\gtrsim 6$ m. All energy of the produced $\gp$ is undetected and the signature is effectively equivalent to the one used for the search $\gp \to \inv$ (invisible mode).
These two signatures closely resemble the ones used in the previous NA64 analysis searching for pseudo-scalar and axion-like particles (see more details in Refs.\,\cite{Banerjee_2019,Banerjee-ALP-2020}). Since the data has
already been unblinded, we chose a conservative approach and re-cast this analysis 
using the pseudo-Dirac model instead of performing a new one. We checked that the optimization of the selection criteria will enhance the sensitivity up to around 10\%. This will be used in the future to perform a blinded analysis when new data will be available. In the following, we provide a brief description of the method.

\noindent Impurities from the beam, consisting mostly in $\pi^-$ ($\lesssim$1\%) are suppressed using the SRD to a level of $\lesssim 10^{-6}$ \cite{Depero:2017mrr}. The tracking system measures the incoming energy of the electrons which is required to be within $\sim$ 3\,GeV from the nominal 100\,GeV beam energy. The  ECAL serves as an active dump measuring the energy deposition of the incoming particles. We require no activity in the VETO and the first HCAL module to reject any initial beam particles that penetrate the ECAL. For S2, no activity is required in all HCAL modules, since $\chi_2$ is assumed to decay outside the fiducial volume. Finally, for S1, a cut $R < 0.06$ is applied on the variable $R \equiv (E_{HCAL} - E^c_{HCAL}) / E_{HCAL}$, defined as the ratio between the energy deposit in the periphery and the total energy deposited in the HCAL2 and HCAL3 modules.

The leading background of S1 is caused by $K^0_L$ traveling undetected through the first HCAL module and deposit energy in either HCAL2 or HCAL3. The S2 signature can instead be misidentified due to large missing energy produced by electron-hadron interactions along the beamline. The total expected background for the two signatures is detailed in the previously published analyses, where it was determined that 0.19$\pm$0.07 events are expected in the signal box of S1~\cite{Banerjee-ALP-2020} and 0.53$\pm$0.17 are expected for S2 \cite{Banerjee_2019, na64-prd}. A full discussion of the uncertainties is also found in these references. In particular, the two leading contributions are $\lesssim 10\%$ for the parametrisation of the form factor in the cross-section and an additional $10\%$ coming from the data-MC discrepancy in the dimuon-yield (events where the interaction $\gamma \to \dimu$ is detected) \cite{na64-prd, Banerjee-ALP-2020}. The difference in the estimate of the $\chidecay$ width, between MadGraph and CalcHEP, leads to a negligible uncertainty $\lesssim 1\%$. All uncertainties, summed in quadrature, do not exceed 20\%.

\noindent The background for this new search is the same as the previous searches for $\ainv$ and $a \to \gamma \gamma$. Thus, the applied cuts are already optimized for the best coverage of $\gp$. An exception is the cut applied on the variable $R$. A larger tail for high values of $R$ is expected in this model, for two reasons: i) smaller $\chi_2$ energy due to the $\chi_1$ emission in the original prompt decay, and ii) the three-body decay of $\chi_2$ which increases the phase space of the final decay products $\chi_1 \ee$. As a result, the efficiency for such events in the S1 signature drops to $\sim$52$\%$ on average, with the exact value being dependent on the parameters of the theory (starting from a minimum value of 45$\%$). 


The signal yield was calculated using a full MC simulation based on the Geant 4 toolkit\cite{geant4}. The framework used for the previous NA64 analysis of the 2016-2018 data \cite{Banerjee_2019, Banerjee-ALP-2020} was extended to include the new model containing the semi-visible decay. Both $\chi_1$ and $\chi_2$ particles are assumed to have no interaction with the detectors. For the S1 signature, $\chi_2$ is forced to decay inside the fiducial volume, i.e., the space between the beginning of HCAL2 and the end of HCAL3. The event is then weighted by the probability of such decay to take place. In the case of S2, we assume that the full energy of $\gp$ is lost, and a weight corresponding to the probability for the Dark Photon to decay beyond all NA64 subdetectors is assigned to the event. This simulation is performed for a grid on the ($m_{\gp}$;$\epsilon$) plane to estimate the expected number of events for different masses and mixing strengths. We use $\alpha_D=0.1$, $m_{\gp} = 3 \cdot m_{\chi_1}$ and a mass splitting $\Delta/m_{\chi_1} = 0.4$ as benchmark for these simulations.



\begin{figure}[tbh!!]
\includegraphics[width=.5\textwidth]{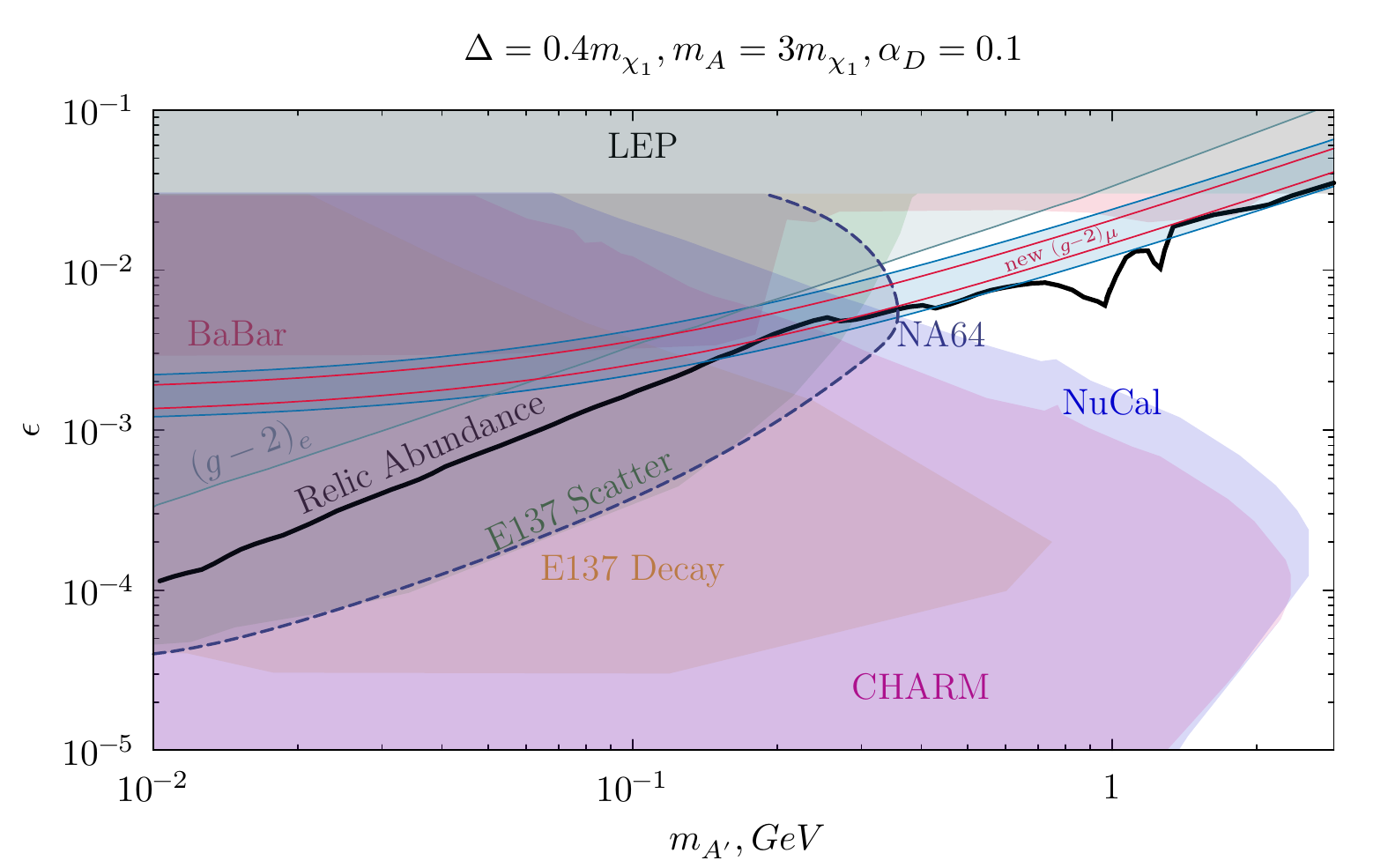}
\vskip-0.cm{\caption{The NA64 90\% exclusion limit for a new vector boson $\gp$ with a coupling to electrons with decay mode $\asemi$. The limits were derived in the ($m_{\gp}$, $\epsilon$) assuming $\alpha_D =0.1$, $m_{\gp}=3\cdot m_{\chi_1}$ and a mass splitting $\bar{f} = 0.4$. The red band shows the region of parameter space within two sigma from the world average of $\gmu$ \cite{PhysRevLett.126.141801}. The blue band shows the same region before the results at Fermilab were published. Constraints from BABAR and E137 are also shown following the recasting done in Ref. \cite{Mohlabeng:2019vrz,Izaguirre:2017bqb}, together with the bounds of NuCal and CHARM \cite{PhysRevLett.126.181801}. A thick black line shows  the combination of parameters compatible with a DM thermal relic scenario. \label{fig:semivis-excl}}}
\end{figure}

\begin{figure}[tbh!!]
\includegraphics[width=.5\textwidth]{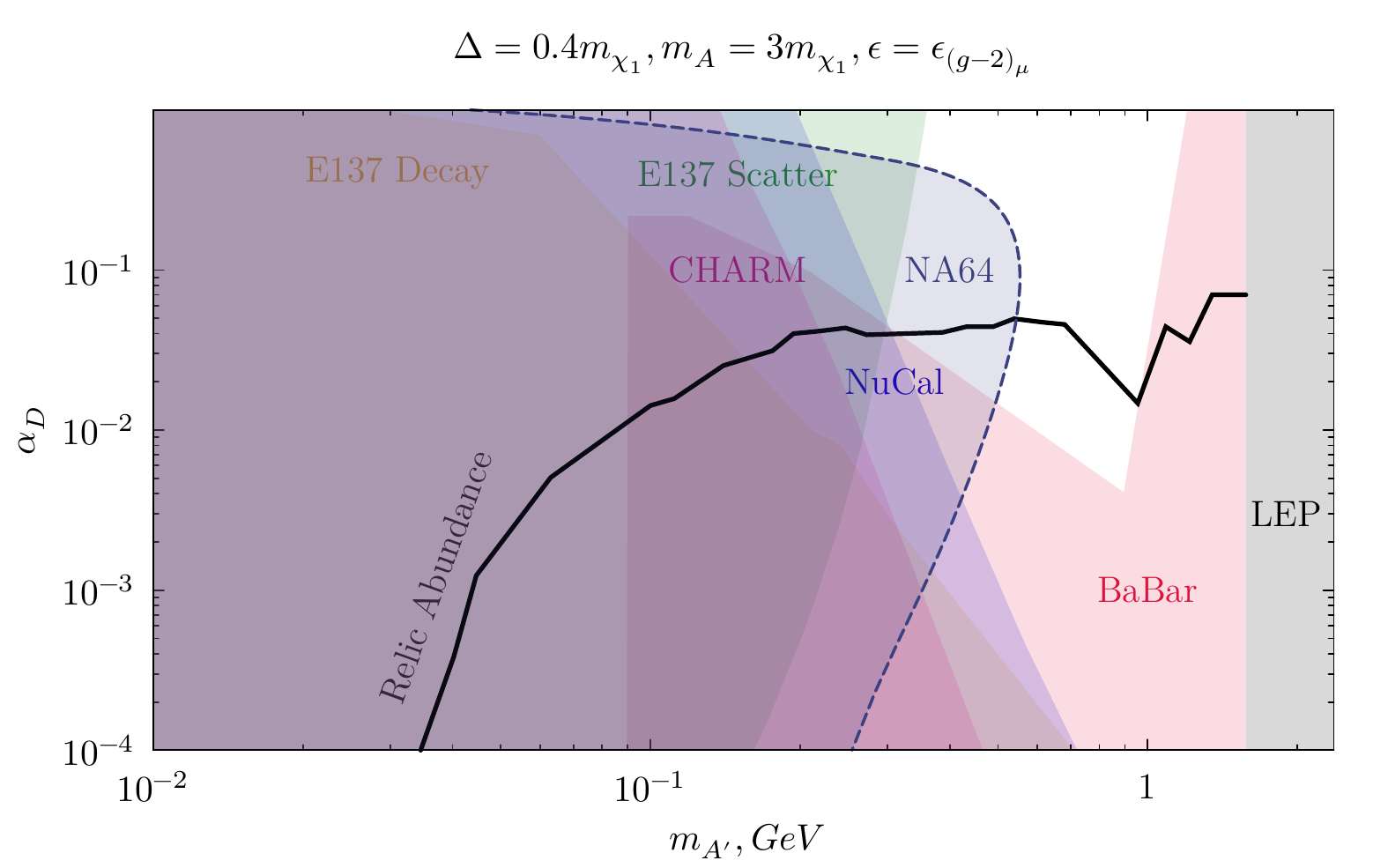} \\
\includegraphics[width=.5\textwidth]{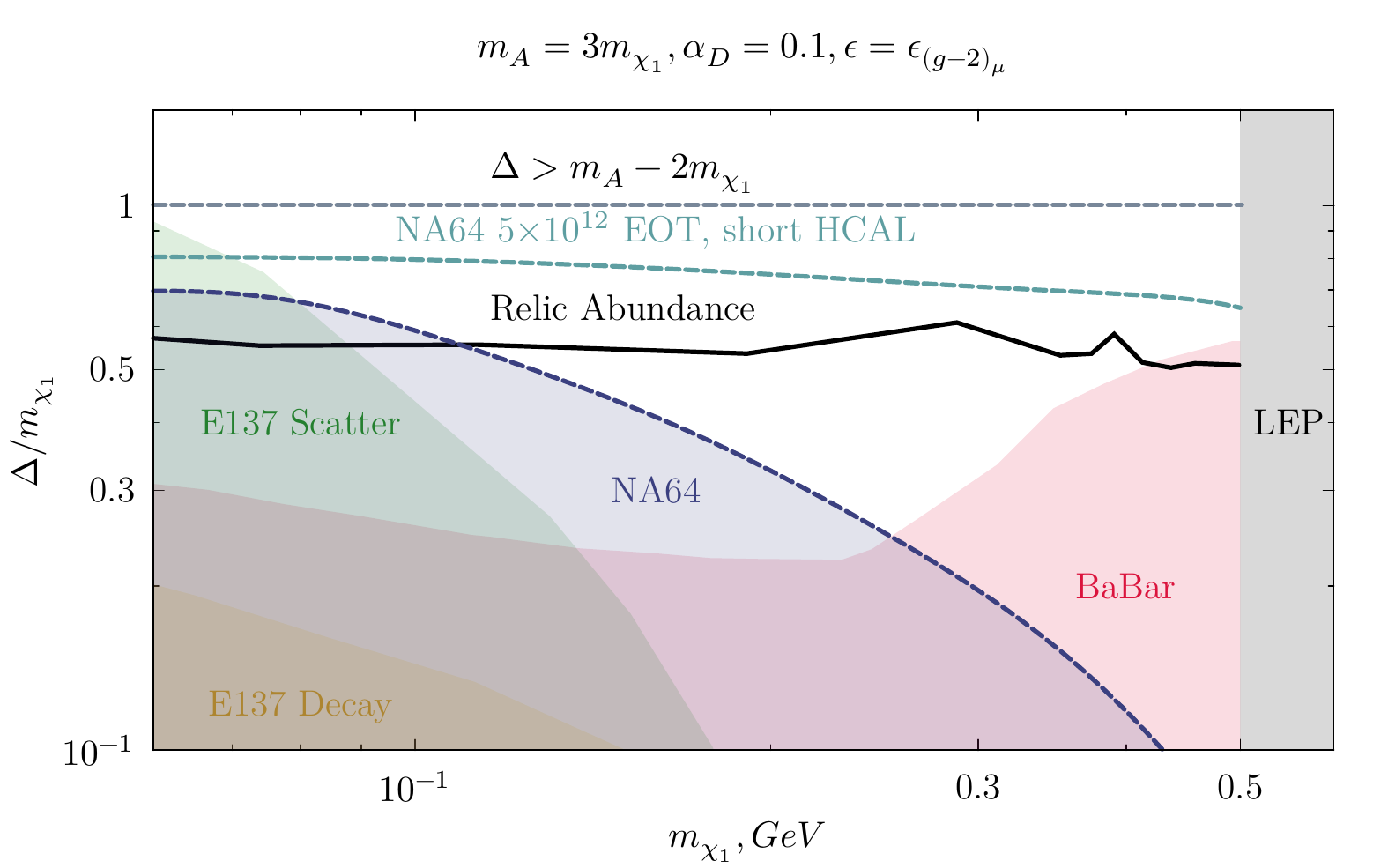} \\
\vskip-0.cm{\caption{The NA64 90\% exclusion limit for a new vector boson $\gp$ with a coupling to electrons with decay mode $\asemi$. The limits were derived in the ($m_{\gp}$;$\alpha_D$) (top) and ($m_{\chi_1}$;$\bar{f}=\Delta / m_{\chi_1}$) (bottom) plane assuming $\alpha_D =0.1$, $m_{\gp}=3\cdot m_{\chi_1}$ and a $\epsilon = \epsilon_{\gmu}$, where $\epsilon_{\gmu}$ is the value in the central band of the $\gmu$ anomaly. Constraints from Babar and E137 are also shown following the recasting done in Ref. \cite{Mohlabeng:2019vrz,Izaguirre:2017bqb}, together with the bounds of NuCal and CHARM \cite{PhysRevLett.126.181801}. A thick black line shows  the combination of parameters compatible with a DM thermal relic scenario. The projected limit for $5\times 10^{12}$ EOT using a compact HCAL1 are drawn in the ($m_{\chi_1}$;$\bar{f}=\Delta / m_{\chi_1}$) plane.\label{fig:semivis-excl-extrapar}}} %
\end{figure}

The exclusion limit was calculated using the multibin limit setting technique with the modified frequentist approach (the code based on the ROOSTATS package \cite{root-roostats}), using the profiled likelihood as a test statistic \cite{ Junk_1999, Read:2002hq}. The corresponding 90\% exclusion limit was obtained using Eq.4  of \cite{Banerjee-ALP-2020} to compute the expected signal yield. The results are summarized in Fig.\ref{fig:semivis-excl} in the 2D plane ($m_{\gp}$;$\epsilon$), where the relevant estimated bound of E137 and Babar are also shown \cite{Mohlabeng:2019vrz,Izaguirre:2017bqb} together with the projection of Nucal \cite{blum,blum1} and CHARM \cite{sg1,PhysRevLett.126.181801}. Using the benchmark values discussed above, our data exclude the $\gmu$ explanation of the semi-visible model up to a mediator mass $m_{\gp} \lesssim 0.39$ GeV including a so-far uncovered area close to this boundary. 
Even though this might look only like as a slight improvement of the current bounds, we would like to stress that these are the first experimental limits in this region of parameter space obtained with a full analysis of the data including all efficiencies and uncertainties.
In Fig.\ref{fig:semivis-excl-extrapar}, the region of the parameter in the central band of the $\gmu$ anomaly is also shown in the two planes ($m_{\gp}$;$\alpha_D$) and ($m_{\chi_1}$;$\Delta/m_{\chi_1}$). To project this space we use the same assumption as in Fig.\ref{fig:semivis-excl} for the parameters not considered, and we set $\epsilon = \epsilon_{\gmu}$ as the epsilon compatible with the central band of $\gmu$, a convention used in previous studies of this model \cite{PhysRevLett.126.181801, Mohlabeng:2019vrz}.
Our results exclude the unexplored area for $m_{\gp} \gtrsim 0.3$ GeV in the $\gmu$ band, leaving space for models in which  $m_{\gp}$ is larger than 0.4 GeV, or with large mass splitting $\Delta$ can still explain the anomaly. The largest limitation to probe the missing region comes from the increasingly short decay time of $\chi_2$, which makes the chance of detection vanishingly small. This is particularly relevant for the mass splitting $\Delta$ (see Fig.\ref{fig:semivis-excl-extrapar} bottom), since the $\Delta^5$ scaling of the decay width adds a large suppression to the signal yield. For $\Delta / m_{\chi_1} \gtrsim 0.5$, our limits become rapidly weak.
In the current design of the experiment, increasing the number of EOT collected will improve our coverage of the model mostly in the region where the signature S2 dominates. In the $\gmu$ band, both S1 and S2 have a sizeable effect (with S1 dominating for $m_{\gp} \gtrsim 0.4$ GeV), which causes the exponential suppression of the signal yield. For $10^{13}$ EOT, masses $m_{\gp} < 0.5$ GeV can be excluded as explanation of the $\gmu$. To improve this limit, a compact dump and a shorter HCAL1 has to be used for the detection of short-lived $\chi_2$. For a compact HCAL1 of 50 cm, a re-casting of the limits shows that the experiment becomes limited only by very high values of $\Delta / m_{\chi_1}$, justified by the large scaling of this parameter with the decay width (see Eq.\ref{eq:width}). In this scenario, the interesting parameter space that justifies both $\gmu$ and the relic density in a freeze-out scenario is completely covered in all the dimensions of the model, which leaves a very exciting prospect for future searches. The compatibility of such a setup with the one to search for $\gp \to \inv $ is currently being studied. Such a setup would also be beneficial to improve the sensitivity of NA64 to higher couplings of axion-like or pseudo-scalar particles $a \to \gamma \gamma$. %

In this work, we analysed the data collected by the NA64
experiment during three different runs in the “invisible-mode” configuration considering a new pseudo-Dirac scenario characterized by the decay $\asemi$ as signal candidate. In this model, the decay of the mediator $\gp$ results in both SM and DM particles in the final states, for an effective signature that combines features of both invisible and visible mode. This scenario can provide an explanation to the $\gmu$ anomaly, recently confirmed at Fermilab \cite{PhysRevLett.126.141801}, and at the same time is compatible with a freeze-out scenario capable to explain the observed DM relic-density.
The previous limits on this model were improved by this analysis, excluding $m_{\gp} \lesssim 0.39$ GeV at 90\% confidence level (C.L.), assuming a DM coupling $\alpha_D=0.1$ and a mass splitting $\Delta / m_{\chi_1} = 0.4$. A large region of parameter space characterized by short living $\chi_2$ remains unexplored as an exciting prospect for future searches.

\par We gratefully acknowledge the support of the CERN management and staff and the technical staff of the participating institutions for their vital contributions. We are thankful to Andrea Celentano for his carefully reading of the paper and very useful comments. We would like to thank Gunar Schnell for discussions on the possibility to search for semi-visible decays in NA64. This work was supported by the Helmholtz-Institut für Strahlen und Kern-physik (HISKP), University of Bonn, the Carl Zeiss Foundation 0653-2.8/581/2, and Verbundprojekt-05A17VTA-CRESST-XENON (Germany), Joint Institute for Nuclear Research (JINR) (Dubna), the Ministry of Science and Higher Education (MSHE) and RAS (Russia), ETH Zurich and SNSF Grant No. 169133, 186181, 186158 and 197346 (Switzerland), FONDECYT Grants \sloppy{No.1191103}, No. 190845 and No. 3170852, UTFSM PI M 18 13, ANID PIA/APOYO AFB180002 (Chile), Millennium Institute for Subatomic Physics at the High-energy frontier (SAPHIR), ICN2019\_044, ANID, Chile.

\bibliography{../Bibliography/bibliographyNA64.bib,../Bibliography/bibliographyNA64exp.bib,../Bibliography/bibliographyOther.bib}
\bibliographystyle{../Bibliography/na64-epjc}

\end{document}